# EL FENÓMENO DE MADURACIÓN DE OSTWALD. PREDICCIONES DE LAS SIMULACIONES DE ESTABILIDAD DE EMULSIONES SOBRE LA EVOLUCIÓN DEL RADIO CÚBICO PROMEDIO DE UNA DISPERSIÓN

German Urbina-Villalba

Instituto Venezolano de Investigaciones Científicas (IVIC), Centro de Estudios Interdisciplinarios de la Física, Lab. de Fisicoquímica de Coloides, Apartado 20632, Edo. Miranda, Venezuela. Email: guv@ivic.gob.ve

**Resumen**

Se describe el algoritmo de simulaciones de estabilidad de emulsiones (SEE) necesario para estudiar la evolución de una dispersión dodecano/agua sometida a floculación, coalescencia, y maduración de Ostwald. Se muestran los resultados de varias simulaciones en las que se varía el potencial de interacción entre gotas de manera sistemática a objeto de estudiar la influencia que tiene: la carga superficial, la deformabilidad de las gotas, la distribución inicial de tamaño de partículas y las fuerzas de hidratación, en la evolución del sistema. Los estudios arrojan varias conclusiones no triviales acerca de la influencia de la maduración de Ostwald en la variación del radio cúbico promedio de las dispersiones alcano/agua.

**Palabras clave**

Ostwald, Simulaciones, Emulsiones, Estabilidad, Dispersión, Alcano.

## 1. INTRODUCCIÓN

De acuerdo a la ecuación de Laplace, la presión a la que están sometidas las moléculas de una gota ($P_{int}$) difiere de la presión del líquido circundante ($P_{ext}$) en una magnitud que es directamente proporcional a la tensión interfacial de la gota ($\gamma$) e inversamente proporcional a su radio ($R_i$):

$$\Delta P = P_{ext} - P_{int} = 2\gamma / R_i \tag{1}$$

En consecuencia, cuando dos gotas de aceite de distinto tamaño entran en contacto en una emulsión aceite/agua, la más grande "devora" a la más pequeña (el fluido se mueve de mayor a menor presión). Tal comportamiento es bien conocido. Sin embargo, mucho menos trivial resulta el hecho de que la misma diferencia de presión origina un potencial químico distinto entre las moléculas que componen ambas gotas. Como la tensión interfacial es una energía libre por unidad de área, las moléculas de una gota tienen en promedio una energía libre $4\pi R_i^2 \gamma / N_{m,i}$ veces superior a las moléculas de aceite puro (donde $N_{m,i}$ es el número de moléculas de la gota $i$ [Acosta, 2003]). Dado que el número de moléculas de una gota esférica es igual a: $(4/3)\pi R_i^3 \rho_i / PM$ (donde $\rho_i$ y $PM$ son la densidad de la fase interna de la gota y el peso molecular de su componente, respectivamente) se estima que la referida diferencia de potencial químico es inversamente proporcional al radio de la gota. De allí que en su camino al equilibrio y en ausencia de contacto directo, las gotas intercambien moléculas de aceite a través del solvente de manera espontánea. Tal fenómeno se conoce con el nombre Maduración de Ostwald.

## 2. ANTECEDENTES

### 2.1 Teoría de LSW

La teoría más importante de Maduración de Ostwald es la teoría LSW formulada por dos autores rusos y uno alemán de manera independiente hace apenas 50 años [Liftshitz, 1961; Wagner, 1961]. Dicha teoría parte de las siguientes premisas:

a) Las partículas suspendidas están fijas en el espacio.

b) El sistema es infinitamente diluido por lo que no existen colisiones entre partículas.

c) La concentración de fase interna (aceite en nuestro caso) en el líquido circundante (agua) es la misma a través de todo el sistema excepto en una vecindad de las gotas del orden de su radio. En esa región la concentración de aceite C es igual a:

$$C(R_i) = C_\infty \exp(\alpha / R_i) \tag{2}$$

Donde $C_\infty$ es la solubilidad a la que estamos acostumbrados: solubilidad del aceite en agua *en presencia de una interfase plana*, y $\alpha$ es el llamado Radio Capilar:

$$\alpha = 2\gamma V_{molar} / RT \tag{3}$$

En la ecuación (3), $V_{molar}$ es el volumen molar del aceite, $R$ la constante universal de los gases y $T$ la temperatura absoluta. Cabe resaltar que $\alpha$ es generalmente del orden de los nanómetros. En consecuencia y de acuerdo a la ecuación (2), la solubilidad en la vecindad de las gotas sólo es sensiblemente distinta a $C_\infty$ cuando el radio de las partículas es del orden de los nanómetros. Más aún, la solubilidad de las partículas disminuye a medida que su radio aumenta.

Bajo las premisas anteriores y para el llamado Régimen Estacionario que corresponde a un tiempo muy largo inespecífico, se deduce que:

d) En cualquier instante del proceso de maduración existe un radio crítico ($R_c$). Este radio corresponde al radio promedio de la distribución de tamaño pesada *en número de partículas* [Finsy, 2004]. Es decir:

$$R_c = \langle R \rangle = \sum R_i / N_p \tag{4}$$

Donde $N_p$ es el número total de partículas.

e) El radio crítico evoluciona en el tiempo a medida que las gotas de la emulsión cambian de tamaño. Partículas con radios menores al crítico se disuelven, mientras que partículas con radios superiores crecen.

f) A tiempos muy largos, la distribución de tamaños adquiere una forma característica conocida como LSW. Esta distribución asemeja la imagen especular de una distribución log-normal con una variación asintótica (cola) que apunta hacia menores radios de partículas. Así, contrario a nuestra intuición inicial, el resultado del proceso de intercambio no es una distribución monomodal de gotas grandes. Más aún, si graficamos la distribución LSW empleando el radio escalado de las partículas $u = R_i / R_c$ se predice que la distribución se hace "estacionaria" (solapa sobre sí misma), y para $0 \leq u \leq 1.5$ corresponde a la siguiente ecuación analítica:

$$W(u) = \frac{81\,e\,u^2\,\exp\left(1/\left([2u/3]-1\right)\right)}{\sqrt[3]{32}\,(u+3)^{7/2}\,(1.5-u)^{11/3}} \tag{5}$$

Donde e es la base del logaritmo neperiano, W(u) es la frecuencia de partículas de tamaño u, y W(u) = 0 para u > 1.5.

g) La velocidad del proceso definida como la variación temporal del radio cúbico promedio, es constante e igual a:

$$V_{LSW} = \frac{dR_c^3}{dt} = \frac{4\alpha D_m C_\infty}{9} \tag{6}$$

Donde $D_m$ es la constante de difusión de las moléculas de aceite en agua.

h) El número de partículas por unidad de volumen disminuye con el tiempo según:

$$n = \left[\frac{1}{2\alpha D_m C_\infty}\right]\frac{1}{t} \tag{7}$$

Cuando se contrastan las fórmulas descritas con la evidencia experimental se hallan resultados mixtos [Weers, 1999]. Aunque existen sistemas en los cuales se ha observado la distribución LSW [Schmitt, 2004], generalmente las distribuciones de tamaño muestran colas hacia la derecha (hacia mayores tamaños de partícula). A veces se observan una ó dos regiones de variación lineal de $V_{LSW}$ vs. $t$, o curvas no-monótonas en vez de rectas. También es muy común que los valores de $V_{LSW}$ difieran en órdenes de magnitud de las velocidades observadas $V_{obs}$:

$$V_{obs} = \frac{dR_p^3}{dt} \tag{8}$$

Donde $R_p$ es el radio promedio de la emulsión bien sea en número, volumen ó área.

Las emulsiones alcano/agua en ausencia de surfactante se encuentran entre los sistemas más simples que permiten contrastar las predicciones teóricas de LSW con el experimento. La ausencia de surfactante evita las complicaciones que podrían surgir de un proceso mixto de transferencia de fase interna en el cual participen las micelas. Por ello resulta sorprendente que para emulsiones de alcano en agua [Sakai *et al.*, 2002] que incluyen ciclohexa-no, n-hexano, n-octano, n-decano, n-tetradecano y n-hexadecano las tasas de $V_{obs}$ sean 32, 97, 679, 8, 3 y 17 veces las predichas por LSW ($V_{LSW}$).

Las emulsiones anteriormente referidas fueron preparadas por sonicación a 40 kHz de 2-8 minutos. El radio promedio de gota se determinó a través de medidas de dispersión de luz. Se encontró que a los 4 minutos los alcanos con un número de carbonos mayor ó igual a diez, mostraban distribuciones con colas a la derecha (hacia mayores tamaños de partícula) y en algunos casos inclusive se observaba un pequeño pico adicional. Se hizo evidente que tal variación era contraria a LSW y cónsona con un proceso de floculación y coalescencia. Los alcanos de menor peso molecular mostraban distribuciones anchas y de muchos picos.

Sakai *et al.* (2001) emplearon microscopía electrónica de crio-fractura para estudiar el comportamiento de nanogotas de aceite en agua. Dichas gotas exhibían un potencial superficial de -35 mV en ausencia de surfactante. Justo después de la sonicación se observaban diámetros entre 30 y 100 nm, y agregados de tamaño mediano entre 200 y 500 nm compuestos de gotas pequeñas. Una hora después se observó que las pequeñas gotas floculadas habían coalescido dando origen a gotas de mayor tamaño. De hecho, éstas gotas de tamaño mediano también coalescían formando gotas de tamaño superior a 1 micra. Sakai *et al.* (2003) elaboraron un modelo basado en el camino libre medio entre gotas según el cual la coalescencia ocurría dentro de los agregados luego de la floculación. Las gotas resultantes (secundarias) volvían a agregarse y a coalescer entre ellas generando una nueva generación de gotas y así sucesivamente. De acuerdo a este modelo y en concordancia con los experimentos, los radios resultantes guardaban una serie geométrica con un común denominador: $(3\phi)^{1/3}$. Esos resultados sugieren que para emulsiones sin surfactante la coalescencia no ocurre durante la colisión entre las gotas sino en un tiempo posterior. Sorprendentemente el modelo sugiere que sólo las gotas de cada generación se agregan entre ellas.

Con los años han surgido varias modificaciones de la teoría LSW que intentan explicar algunas de sus inconsistencias. Generalmente se supone que la tasa de maduración debe ir multiplicada por un factor que depende de la fracción de volumen para tomar en cuenta el hecho de que las gotas no están fijas en el espacio. Otras teorías han intentado establecer el rol de los surfactantes en el proceso de maduración. En este sentido los resultados son altamente contradictorios. Por ejemplo, Hoang *et al.* (2004) encontró que el incrementar la concentración de surfactante iónico en una dispersión aceite/agua no tiene un efecto sobre $V_{obs}$, mientras que el añadir un surfactante no-iónico la incrementa sensiblemente. Algunos autores afirman que las micelas deben jugar un papel fundamental en la transferencia de aceite entre gotas, pero los estimados teóricos y algunas medidas experimentales contradicen esta versión [Kabalnov, 1994; Kabalnov, 1996]. Kabalnov *et al.* [De Smet, 1997; Kabalnov, 1992] han sugerido que los surfactantes forman una especie de membrana alrededor de las gotas que obstaculiza la transferencia de la fase interna al agua haciéndola dependiente de la solubilidad del aceite en la membrana. De acuerdo a éstas estimaciones, la distribución resultante tiende a ser más acampanada y simétrica, y los valores de $V_{OR}$ menores.

Para macro-emulsiones concentradas las medidas de Schmitt y colaboradores [Schmitt, 2004] sugieren que la maduración de Ostwald y la coalescencia ocurren de manera secuencial, y por tanto prevalecen a tiempos distintos. Estas variaciones se caracterizan por índices de polidispersidad bajos y altos, respectivamente. En consecuencia, existe un tiempo crítico en el cual las tasas de Ostwald y coalescencia se hacen iguales. Dado que la coalescencia depende del número de rupturas de las películas formadas entre las interfases de las gotas agregadas, ella es proporcional al área total de las gotas de la emulsión. Por otra parte, la velocidad de maduración es proporcional al radio cúbico de las gotas, y en consecuencia, si las tasas se igualan en algún momento, es posible deducir una expresión analítica para la frecuencia de coalescencia y evaluarla experimentalmente.

## 2.2 Evaluación de la tasa de desestabilización de una emulsión basada en la variación del número de partículas por unidad de volumen.

En general es difícil determinar si el tamaño de gota se incrementa por maduración de Ostwald ó a través de un proceso mixto de floculación y coalescencia. Se conoce que la coalescencia aumenta la polidispersidad de una dispersión. En consecuencia puede parecer difícil obtener una variación lineal de $R^3$ vs. $t$ producto de floculación y coalescencia.

Sin embargo, la variación del número de total de gotas por unidad de volumen para el caso de maduración de Ostwald sigue una ecuación diferencial cuadrática similar a la del fenómeno de floculación. Esto se hace evidente relacionando el radio promedio de las partículas de una dispersión con el número de partículas por unidad de volumen. Así:

$$\frac{4}{3}\pi R_i^3 N_p / (V_O + V_W) = \frac{4}{3}\pi R_i^3 n = \phi \qquad (9)$$

Donde $\phi$ es la fracción de volumen de aceite en agua. Así:

$$V_{LSW} = \frac{dR_c^3}{dt} = \frac{3\phi}{4\pi}\frac{d}{dt}\left(\frac{1}{n}\right) = \frac{3\phi}{4\pi}\left(-\frac{1}{n^2}\right)\frac{dn}{dt} =$$

$$= \frac{4\alpha D_m C_\infty}{9} \qquad (10)$$

Por lo que:

$$\frac{dn}{dt} = -\left[\frac{4\alpha D_m C_\infty}{9}\right]\left[\frac{4\pi}{3\phi}\right]n^2 \qquad (11)$$

Esta es una ecuación de segundo orden similar a la ecuación de Smoluchowski (1917) para el número total de partículas que floculan de manera irreversible por unidad de volumen:

$$\frac{dn}{dt} = -\left[4\pi D R_i\right]n^2 \qquad (12)$$

Donde D es el coeficiente de difusión de las gotas. En consecuencia la ecuación (11) debe tener el mismo resultado de Smoluchowski para la variación de n excepto por el valor de la tasa de agregación ($k_F$) que queda ahora sustituida por $k_O$:

$$n_g = \frac{n_0}{1 + k_O n_0 t} \qquad (13)$$

Donde $n_0 = n(t = 0) = n_g(t = 0)$, $n_g$ es el número de gotas, y $k_O$ es igual a $(4\pi/3\phi)$ veces la tasa de maduración $V_{LSW}$:

$$k_O = \frac{16\pi\alpha D_m C_\infty}{27\phi} \qquad (14)$$

La ecuación (14) debe reproducirse si la maduración de Ostwald no produce la disolución total de las partículas pequeñas. Esto sólo puede ocurrir si la tasa de floculación y coalescencia es más rápida que la de maduración. En consecuencia, discriminar si el incremento del radio promedio de partículas de una dispersión es producto del mecanismo de maduración de Ostwald ó del fenómeno mixto de floculación y coalescencia es difícil. En el caso más común, todas ésas contribuciones están presentes e incluidas en el valor de una tasa de desestabilización mixta ($k_{FCO}$).

### 2.2.1 Determinación de la constante de floculación de suspensiones empleando medidas de turbidez.

Para el caso de la floculación irreversible de partículas sólidas, Smoluchowski (1917) dedujo que el número de agregados de tamaño k (formado k partículas primarias) presente en la dispersión en el tiempo t es igual a:

$$n_k = \frac{n_0 (k_F n_0 t)^{k-1}}{(1 + k_F n_0 t)^{k+1}} \qquad (15)$$

De acuerdo al esquema propuesto por Smoluchowski, el proceso de agregación se inicia con la formación de dobletes:

$$\frac{dn_1}{dt} = -k_{11} n_1^2 - k_{12} n_1 n_2 - \dots \qquad (16)$$

$$\frac{dn_2}{dt} = \frac{k_{11}}{2} n_1^2 - k_{12} n_1 n_2 - \dots \qquad (17)$$

Donde $n_1$ y $n_2$ son las concentraciones de singletes y dobletes.

La turbidez de una suspensión puede definirse como [Lips, 1973]:

$$\tau = \sum_{k=1}^{\infty} n_k (t, k_F) \sigma_k \qquad (18)$$

Donde $\sigma_k$ es la sección transversal óptica de un agregado de tamaño k. Diferenciando la Ec. (18) se obtiene:

$$\frac{d\tau}{dt} = \sigma_1 \frac{dn_1}{dt} + \sigma_2 \frac{dn_2}{dt} + \dots \qquad (19)$$

A tiempos muy cortos ($t \rightarrow 0$), $n_2$ se aproxima a cero y por tanto los términos superiores de las ecuaciones (16)-(18) pueden ser despreciados. Así se obtiene que:

$$\left(\frac{d\tau}{dt}\right)_0 = 230\left(\frac{dAbs}{dt}\right)_0 = \left(\frac{1}{2}\sigma_2 - \sigma_1\right)k_{11} n_0^2 \qquad (20)$$

Aquí, $\sigma_1$ y $\sigma_2$ son las secciones transversales ópticas de una partícula esférica y de un doblete, $\tau = (1/L) \ln (I_0/I)$, Abs $= log(I_0/I)$ es la absorbancia de la dispersión, $I$ es la intensidad de la luz emergente de la celda de medida de ancho $L$ (generalmente $10^{-2}$ m), e $I_0$ es la intensidad de la luz incidente.

Las secciones transversales ópticas pueden ser evaluadas empleando la teoría de Rayleigh, Gans, y Debye (RGD) [Kerker, 1969]. Así la constante de formación de dobletes ($k_{11}$) puede ser obtenida mediante medidas de turbidez.

La teoría RGD sólo es válida cuando:

$$C_{RGD} = (4\pi R_l / \lambda)(m-1) << 1 \qquad (21)$$

Donde $\lambda$ es la longitud de onda de la luz en el medio y $m$ es el índice de refracción relativo entre la partícula y el medio. Esto permite evaluar numéricamente expresiones del tipo:

$$\sigma_k = \sigma_{k,a} =$$
$$= \frac{4}{9}\pi R^2 \alpha^4 (m-1)^2 \int_0^{\pi} P_k(\vartheta)(1+\cos^2\vartheta)\sin(\vartheta)\,d\vartheta \qquad (22)$$

Donde $\vartheta$ es el ángulo de dispersión de luz, $\alpha = 2\pi R_l/\lambda$, y $P_k(\vartheta)$ es el factor de forma de un agregado de tamaño $k$ deducido por Puertas *et al.* (1998) y Puertas y de las Nieves (1997). Obsérvese que la sección transversal depende del radio de las partículas que componen el agregado, el cual entra en la expresión en calidad de parámetro fijo, el cual se introduce como dato inicial en el cálculo.

La expresión (22) puede evaluarse numéricamente. En consecuencia el empleo de la ecuación (15) para determinar el número de agregados de cada tamaño por unidad de volumen existentes en el tiempo t, permite evaluar la turbidez (Ec. (18)). Para esto es necesario utilizar $k_F$ como parámetro de ajuste entre la data experimental de turbidez y los valores teóricos predichos por la ecuación (18).

### 2.2.2 Determinación de la constante mixta de floculación y coalescencia empleando medidas de turbidez.

Recientemente Rahn-Chique *et al.* [Rahn-Chique, 2012a; 2012b; 2012c] formularon un modelo que toma en cuenta la ocurrencia simultánea de floculación y coalescencia en una emulsión. En este caso la turbidez resulta de varias contribuciones que incluyen las gotas primarias de la dispersión, agregados de gotas primarias, gotas esféricas más grandes resultantes de la coalescencia de gotas pequeñas, y agregados "mixtos" que se originan tanto por la floculación de las gotas grandes con las pequeñas, como de la coalescencia parcial de los agregados de gotas peque-ñas. Si el número de agregados "mixtos" es despreciable, puede deducirse que la turbidez depende de la fracción ($g$) del número total de agregados ($n$) que resulta de la agregación irreversible de las gotas primarias existentes (inicialmente) en la emulsión:

$$\tau = n_1(k_F,t)\sigma_1 + g\sum_{k=2}^{k_{max}} n_k(k_F,t)\sigma_{k,a} + $$
$$\qquad\qquad\qquad\qquad (23)$$
$$+ (1-g)\sum_{k=2}^{k_{max}} n_k(k_F,t)\sigma_{k,s}$$

En consecuencia ($1-g$) constituye la fracción de agregados que son producto de la coalescencia de las gotas. Cabe destacar que en la teoría de Smoluchowski, un agregado de $k$ partículas contiene $k$ partículas primarias. Por tanto, su volumen es igual a $k$ veces el volumen de una partícula primaria individual ($v_l$). De esta manera, si el proceso de coalescencia no ocurre durante la floculación de las gotas sino un tiempo después (entre algunas de las gotas que conforman el agregado), igualmente se producirá un agregado de volumen $k\,v_l$. Por lo tanto, la fracción ($1-g$) resulta tanto de la coalescencia instantánea de gotas al contacto, como de su coalescencia "retardada" dentro de un agregado de partículas primarias previamente formado.

Nótese que las dos últimas sumatorias de la Eq. (23) puede agruparse, sacando como factor común $n_k$. Tal rearreglo conlleva a una expresión en la que "$g$" funge tan sólo como un parámetro ajustable de una sección transversal efectiva, que mide más bien el grado de no-esfericidad de los agregados formados.

La sección transversal $\sigma_{k,s}$ corresponde a esferas de mayor tamaño que las gotas inicialmente presentes en la emulsión. Esta sección puede deducirse notando que en el caso en que ocurriese la coalescencia de todas las partículas de un flóculo, el radio resultante de la nueva gota (agregado esférico de mayor tamaño) sería igual a: $R_k = \sqrt[3]{k}\,R_0$. Así pues:

$$P_k(\vartheta) = P_{k,s}(\vartheta) = \left(3\frac{sin\,u_k - u_k\,cos\,u_k}{u_k^3}\right)^2 \qquad (24)$$

Donde $\alpha_k = 2\pi R_k/\lambda$, y $u_k = 2\alpha_k \sin(\vartheta/2)$.

Al igual que en el caso de suspensiones la constante $k_F$ se determina por ajuste de la turbidez experimental a la ecuación (23). Sin embargo en este caso la variable "$g$" también se ajusta. Para suspensiones ó emulsiones estabilizadas con surfactantes iónicos la agregación se promueve aumentando la fuerza iónica (FI) del medio (concentración de electrolitos en la fase externa), dado que esto origina el apantallamiento de las cargas electrostáticas entre las partículas y su subsiguiente agregación.

En trabajos anteriores [Urbina-Villalba, 2004; 2005; 2006; 2009; Osorio, 2011] hemos demostrado, que si las gotas son no-deformables y el potencial repulsivo entre las mismas está apantallado, la constante de floculación incluye el efecto de la coalescencia. En ese caso la constante que se obtiene es $k_{FC}$:

$$n = \frac{n_0}{1 + k_{FC} \, n_0 \, t} \quad (25)$$

Dada la forma de las ecuaciones (13) y (25) es claro que en el caso de que la maduración de Ostwald ocurra, su efecto queda incorporado en la determinación de la tasa de desestabilización ($k_{FCO}$). Sin embargo, debe observarse también que si bien el proceso de maduración de Ostwald produce gotas esféricas al igual que el proceso de coalescencia, el incremento de su tamaño no es consistente ni con la ecuación (15) ni con a la expresión $R_k = \sqrt[3]{k} R_0$, que es la que permite evaluar la sección transversal óptica a partir de la ecuación (24). En consecuencia, las ecuaciones anteriores sólo permiten establecer la influencia del proceso de maduración de Ostwald sobre la tasa de agregación y coalescencia, más no la tasa de ocurrencia del proceso de Ostwald únicamente. Se necesitaría para ello deducir nuevas expresiones para $R_k$ y $n_k$.

### 2.3 Evaluación de la tasa de desestabilización de una emulsión basada en la variación del radio cúbico promedio.

La mayoría de los equipos de dispersión de luz no distingue entre una partícula y un agregado de partículas con el mismo radio hidrodinámico. De hecho, los instrumentos generalmente aproximan los agregados de partículas por esferas con un radio efectivo conveniente. Es así como pueden estimar la sección transversal óptica ($\sigma_i$) de las partículas que les es necesaria para calcular su tamaño. Por ello es difícil establecer el origen de la variación del radio promedio de una emulsión.

En el caso de un sistema de gotas que se agregan, el cálculo del radio promedio necesita el número de partículas por unidad de volumen de cada tamaño:

$$\langle R \rangle = \sum_i R_i \, n_i \quad (26)$$

Si las condiciones fisicoquímicas son tales que sólo se forman agregados lineales: $R_i = i \, R_0$ y:

$$\langle R \rangle = R_0 \, (1 + k_F \, n_0 \, t) \quad (27)$$

Entonces:

$$V_F = \frac{d \langle R \rangle^3}{dt} = 3 \, k_F \, n_0 \, R_0^3 \, (1 + k_F \, n_0 \, t)^2 \quad (28)$$

Esta ecuación presenta como límite a tiempos pequeños: $3 \, k_F \, n_0 \, R_0^3$, lo cual permite evaluar el valor de $k_F$ a partir del radio promedio. En los casos más generales de agregación controlada por difusión (*Difussion Limited Cluster Aggregation, DLCA*):

$$\langle R \rangle = R_0 \, (1 + k_F \, n_0 \, t)^{1/d} \quad (29)$$

Donde: $d$ es el coeficiente fractal del agregado. Inclusive en el caso de agregación lenta (*Reaction Limited Cluster Aggregation, RLCA*) para el cual:

$$\langle R \rangle = R_0 \, \exp(k_F \, n_0 \, t / d) \quad (30)$$

El procedimiento descrito arriba (elevar el radio al cubo, derivar respecto al tiempo y calcular el límite a tiempos cortos) lleva a un valor límite similar:

$$V_F = \left( \frac{d \langle R \rangle^3}{dt} \right)_{t \to 0} = (3/d) \, k_F \, n_0 \, R_0^3 \quad (31)$$

Esta ecuación permite determinar $k_F$ a partir de la pendiente inicial del radio cúbico promedio en función del tiempo.

## 3. DESARROLLO TEÓRICO

En este artículo se emplean simulaciones de estabilidad de emulsiones para estudiar la variación de radio cúbico promedio de una emulsión dodecano/agua sin surfactante durante los primeros minutos de evolución del sistema (t < 5 min) luego de su preparación:

$$V_{Teórico} = \frac{d \langle R \rangle^3}{dt} \quad (32)$$

Se pretende contrastar los resultados de las simulaciones con las tasas de maduración de Ostwald y las distribuciones de tamaño de partícula obtenidas por Sakai *et al.* (2002) para el mismo sistema a 4 minutos de evolución.

### 3.1 Simulaciones de Estabilidad de Emulsiones

Una descripción detallada del algoritmo de SEE puede se encontrada en las siguientes referencias [Urbina-Villalba, 2000; 2003; 2009a; Toro-Mendoza, 2010; Osorio, 2011]. Aquí solo se ilustran aquellos aspectos esenciales a las simulaciones que se discuten.

Las simulaciones comienzan por distribuir $N_p$ gotas de manera aleatoria en una caja cúbica de arista $L$. Las partículas se mueven de un modo similar a como lo hacen en las simulaciones de Dinámica Browniana [Ermak, 1978]. El desplazamiento de la partícula $i$ en el tiempo $\Delta t$: $\vec{r}_i(t + \Delta t) - \vec{r}_i(t)$, es el resultado de dos tipos de fuerzas:

1) Las fuerzas deterministas que incluyen las fuerzas entre partículas $\sum_{i=1; j \neq i}^{N} \vec{F}_{ji}$ y las resultantes de su interacción con un campo externo $\vec{F}_{ext}$.

2) Las fuerzas aleatorias que surgen del intercambio térmico de momento entre las partículas y el solvente. Estas generan el llamado movimiento Browniano. A fin de reproducir las propiedades estadísticas de este movimiento se genera un vector de números aleatorios que cumplen con la característica de formar una distribución Gausiana de media cero y varianza unidad. El desplazamiento cuadrático medio típico del movimiento Browniano [Ermak, 1978] se obtiene multiplicando cada desviación aleatoria por: $\sqrt{2 D_{eff,i}(d_c, \phi) \Delta t}$. Donde $D_{eff,i}(d_c, \phi)$ es la constante de difusión efectiva de la partícula, la cual toma en cuenta las interacciones hidrodinámicas entre gotas, y depende de una distancia característica ($d_c$) y de la fracción de volumen de aceite alrededor de la partícula ($\phi$). Así:

$$\vec{r}_i(t + \Delta t) =$$
$$= \vec{r}_i(t) + \left\{ \left( \sum_{\substack{j=1 \\ j \neq i}}^{N} \vec{F}_{ji} + \vec{F}_{ext} \right) D_{eff,i}(d_c, \phi) \middle/ k_B T \right\} \Delta t$$
$$+ \sqrt{2 D_{eff,i}(d_c, \phi) \Delta t} [\vec{G}auss] \qquad (33)$$

En el caso de gotas "no deformables":

$$D_{eff,i}(d, \phi) = D_0 f_{corr,i} = (k_B T / 6 \pi \eta R_i) f_{corr,i} \qquad (34)$$

Donde $\eta$ es la viscosidad de la fase externa (agua en este caso), $k_B$ es la constante de Boltzmann, $T$ la temperatura absoluta, y $f_{corr,i}$ una corrección hidrodinámica de campo medio que depende de la fracción de volumen de gotas alrededor de la gota considerada, y de la distancia entre la partícula y su vecina más cercana ($d_c$). Si otra partícula (j) se acerca a una distancia menor del radio de la partícula i (cuyo movimiento se calcula): $r_{ij} - R_i - R_j \leq R_i$ (donde $r_{ij} = |\vec{r}_i(t) - \vec{r}_j(t)|$) entonces la fórmula de Honig *et al.* (1971) es empleada para calcular la difusión resultante:

$$f_{corr,i} = (6u^2 + 4u) / (6u^2 + 13u + 2) \qquad (35)$$

Donde: $u = (r_{ij} - R_i - R_j) / R_R$, and $R_R = 1/2(R_i + R_j)$. De otra forma, la fracción de volumen de partículas es utilizada. En este caso:

$$f_{corr,i} = 1.0 - 1.734 \phi + 0.91 \phi^2 \qquad (36)$$

Al comienzo del cálculo el programa crea una distribución de tamaño de partículas que puede ser mono-dispersa, Gaussiana, log-normal ó LSW. Luego distribuye las mismas de manera aleatoria en la caja de simulación, cuyas dimensiones son calculadas externamente con el fin de reproducir la densidad numérica de partículas por unidad de volumen del experimento. Seguidamente se asigna la constante de difusión a cada partícula siguiendo las fórmulas descritas arriba. El programa permite adicionalmente utilizar expresiones de la difusión de una partícula a dilución infinita distintas de la difusión de Stokes (Ec. (34)).

El programa Emulf.f está codificado en Fortran 77 y tiene actualmente tres modos de corrida que incluyen gotas no deformables, gotas deformables, y una variación dinámica del tipo de gotas de acuerdo a rangos de tamaño. La deformación se simula empleando el modelo simplificado de esferas truncadas propuesto por Danov *et al.* (1993). El modelo supone tres regiones de aproximación [Danov 1993; Ivanov, 1999; Denkov, 1995; Petsev, 1995]. La región I corresponde a distancias de acercamiento grandes en que la interacción hidrodinámica entre gotas es pequeña. Aquí las gotas se comportan como no-deformables y se mueven por tanto como esferas sólidas. La región II cubre distancias entre la distancia inicial de deformación ($h = b_0$, donde $h$ es la menor distancia entre las superficies de las gotas) y la obtención del radio máximo de la película entre gotas $0 \leq r_f \leq r_{f\,max}$. Cuando dos gotas se aproximan la fuerza atractiva entre ellas provee la energía necesaria para acercarlas (a pesar del incremento de la fricción generado por el solvente), la cual es mayor a menor distancia entre las superficies de las gotas. En consecuencia las gotas se deforman mediante un mecanismo complejo convirtiéndose finalmente (y *aproximadamente*) en esferas truncadas. Eso produce una película (gota-agua-gota) del tipo aceite-agua-aceite o O/W/O por sus siglas en inglés. Esa película, observada a través del eje que contiene los centros de ambas gotas, tiene forma de un cilindro, cuyos límites son las interfases planas circulares W/O que limitan el cilindro. El radio de la película es el radio del círculo plano, el cual define la parte truncada de cada esfera. Ese radio del *film*, $r_f$, es –en este modelo– igual en ambas esferas floculadas a pesar de su tamaño.

La expresión que se emplea para definir la distancia inicial de deformación se calculó ajustando una ecuación empírica a un cálculo exacto de Petsev *et al.* (1995) entre dos gotas de igual tamaño. Así surgió una expresión que depende del tamaño de las gotas y de su tensión interfacial W/O [Toro-Mendoza, 2010]:

$$b_0 = (1.2932\,x10^8 - 8.6475\,x\,10^{-9} \exp(-R_i/1.8222\,x\,10^{-6}))\,x$$

$$\frac{(3.3253\,x\,10^{-9} + 5.9804\,x\,10^{-9}\,\exp(-\gamma/0.00402))}{(3.3253\,x\,10^{-9} + 5.9804\,x\,10^{-9}\,\exp(-10^{-3}/0.00402))} \tag{37}$$

Cada vez que dos gotas se aproximan, el programa calcula la expresión (37) para cada gota. Como el radio máximo del *film* O/W/O se aproxima por $r_{f\,max} \approx \sqrt{R_i\,b_0}$ y es igual para ambas gotas, la distancia inicial de deformación entre ellas ($b_0$) se escoge como aquél de los dos valores generado por la Ec. (37) que es compatible con la película de menor radio máximo.

Entre las zonas II y III, la distancia $h_0$ permanece constante y el radio de la película evoluciona desde un valor mínimo hasta el valor máximo dado por la Ec. (38):

$$r_{film} = \sqrt{R_i^2 - \{R_i\,(r_{ij} - b_0)/(R_i + R_j)\}^2} \tag{38}$$

Es importante notar que aún cuando la distancia $b_0$ es constante, la distancia entre los centros de las gotas $r_{ij}$ cambia como producto del crecimiento del radio del film. En consecuencia las fuerzas entre las partículas no son nulas en la región II.

Luego de que se alcanza el radio máximo, $r_{f\,max}$ el film comienza a "drenar" ($h$ disminuye a $r_f = r_{f\,max}$ constante). En esta región:

$$h = r_{ij} + \sqrt{R_i^2 - r_{f\,max}^2} + \sqrt{R_j^2 - r_{f\,max}^2} \tag{39}$$

La coalescencia ocurre dentro de la región III cuando se alcanza la distancia crítica $h_c$, la cual puede aproximarse por:

$$h < h_c = (A_H \lambda_c / 128\gamma)^{1/4} \tag{40}$$

Donde: $\lambda_c = r_f/10$. A esta distancia el film se rompe y las gotas coalescen.

En las regiones de esferas truncadas, la constante de difusión se aproxima por la expresión de Danov (1993):

$$D_{Danov} = D_0 f_{corr} =$$
$$= 4 D_0 (h/R_i)(1 + (r_f^2/R_i\,b) + (r_f^4/R_i^2\,b^2)\,\varepsilon_s)^{-1} \tag{41}$$

Donde $\varepsilon_s$ es un parámetro que depende de la movilidad de la capa de surfactante en la interfase.

Las fuerzas de interacción entre partículas se calculan derivando los potenciales de interacción correspondientes:

$$F_{ij} = -\partial V(r_{ij}) / \partial r_{ij} \tag{42}$$

La ecuación de movimiento de gotas deformables y no deformables es la misma (Ec. (33)) y sólo difiere en las expresiones de las fuerzas y los "tensores" de difusión en cada caso.

En estas simulaciones el aceite de las gotas interacciona atractivamente a través de la ecuación de van der Waals. Para gotas esféricas [Hamaker, 1937]:

$$V_A = V_{vdW} =$$
$$= -A_H/12\,(y/(x^2 + xy + x) + y/(x^2 + xy + x + y)$$
$$+ 2\ln[(x^2 + xy + x)/(x^2 + xy + x + y)]) \tag{43}$$

Donde: $x = h/2R_i$, $R_i/R_j$, y $A_H$ es la constante de Hamaker.

Para el caso de gotas deformables la ecuación correspondiente es [Danov, 1993]:

$$
\begin{aligned}
V_{vdW} = & -\frac{A_H}{12}\left(\frac{2R_2\,(l_1-h)}{l_1(l_2+h)} + \frac{2R_2\,(l_1-h)}{h\,(l_1+l_2)} + 2ln\left(\frac{h\,(l_1+l_2)}{l_1\,(l_2+h)}\right) + \frac{r_f}{h^2}\right. \\
& -\frac{l_1-h}{l_2}\left(\frac{2r_f^2}{h\,l_1}\right) - \frac{l_1-R_1-(l_2-R_2)}{2l_1-2R_1-h}\left(\frac{2r_f^2}{h\,l_1}\right) - \frac{2\,(l_2-R_2)-h}{2l_1-2R_1-h}\left(\frac{d-h}{2h}\right) \\
& +\frac{2R_2 l_2^2\,(l_1-h)}{h l_1\,(l_1+l_2)\,(l_2+h)} - \frac{2R_2^2}{h\,(2l_1-2R_1-h)}\,\frac{l_1^2+r_f^2}{(l_2+l_2)\,(l_2+l_2-2R_2)} \\
& +\frac{2R_2^2\,d}{(2l_1-2R_1-h)\,[(l_2+h)\,(h+l_2-2R_2)-(l_1-h)\,(l_1-2R_1-h)\,]} \\
& \left.+\frac{4\,R_2^3(l_1-h)}{(h+l_2)\,(h+l_2-2R_2)-(l_1-h)\,(l_1-2R_1-h)}\,\frac{1}{(l_1+l_2)(l_1+l_2-2R_2)}\right)
\end{aligned}
\tag{44}
$$

Donde: $l_1 = h + R_1 + \sqrt{R_1 - r_f^2}$, $l_2 = h + R_2 + \sqrt{R_2 - r_f^2}$
y $d = \sqrt{h^2 + 4r_f^2}$.

La interacción electrostática entre las cargas existentes en las superficies de las gotas (generadas por adsorción preferencial de oxidrilos) se calcula en ambos caso con la ecuación [Danov, 1993]:

$$
\begin{aligned}
V_E = & \left(64\pi C_{el} k_B T/\kappa\right) \tanh(e\Psi_{0i}/4k_B T)\,\tanh(e\Psi_{0j}/4k_B T)\text{ x} \\
& \exp(-\kappa h)\,[\,r_f^2 + 2R_i R_j/\kappa(R_i+R_j)]
\end{aligned}
\tag{45}
$$

Donde el radio del film es igual a cero en el caso de gotas no deformables, $C_{el}$ es la concentración de electrolito en el medio, $e$ es la unidad de carga electrostática, $\kappa^{-1}$ es la longitud de Debye, y $\Psi_0$ es el potencial superficial eléctrico de las gotas que en ausencia de surfactante es producto de la adsorción de iones oxidrilo [Beattie, 2004; Stachurski, 1996; Marinova, 1996].

En el caso de gotas deformables existen dos potenciales adicionales que deben ser incluidos. La energía de deformación extensional (o "dilatacional") es producto del aumento de área de la gota al deformarse. Esto pone en contacto más moléculas de agua con moléculas de aceite, lo cual origina una contribución interfacial del tipo $\gamma \Delta A$ (donde $\Delta A$ es el cambio de área):

$$
V_d = \gamma \pi r_{f,i}^4 / (2R_i)
\tag{46}
$$

Por otro lado, el potencial de curvatura (*bending*) se origina como producto de la diferencia que existe entre la curvatura espontánea que el surfactante adsorbido quiere adoptar ($H_0$) y la curvatura de la interfase de la gota $H = -1/R_i$:

$$
V_b = -2\pi r_f^2 B_0 H
\tag{47}
$$

Aquí $B_0 = -4 k_b H_0$, y $H_0 = -1/R_{c,0}$, $k_b$ es una constante conocida como el momento de dobladura (*bending* en la teoría de Helfrich). $R_{c,0}$ es la curvatura de la monocapa de surfactante en su configuración de menor energía. En ausencia de surfactante este potencial no debería ser incluido. Sin embargo, la adsorción de OH⁻ a la interfase agua/aceite genera un potencial eléctrico significativo en la interfase. En nuestra opinión, esos radicales actúan como surfactantes por lo que pueden experimentar una energía de frustración por curvatura, similar a la de los surfactantes. En consecuencia, este potencial también fue tomado en cuenta.

Finalmente algunos de los cálculos incluyen un potencial de hidratación que tiene la forma [Ivanov, 1999]:

$$
V_{hid} = \pi R_i \lambda_0 f_0 \exp(-h/\lambda_0)
\tag{48}
$$

Donde los valores de sus parámetros son aproximadamente: $\lambda_0 = 0.6$ nm, y $f_0 = 3$ mJ/m².

Una vez que las constantes de difusión han sido asignadas y las fuerzas calculadas, las gotas se mueven siguiendo la ecuación (33). En el caso de gotas no-deformables la coalescencia ocurre cuando la distancia entre los centros de las gotas se hace menor que la suma de sus radios. En este caso se crea una nueva gota en el centro de masas de las partículas interactuantes ($R_{new} = \sqrt[3]{R_i^3 + R_j^3}$). Lo mismo ocurre en el caso de gotas deformables cuando la separación entre gotas $h$ alcanza la distancia crítica $h_c$ (Ec. 40).

En los cálculos de gotas deformables el programa posee un mecanismo adicional de coalescencia que se basa en la presencia de ondas capilares sobre las superficies de las gotas. Cada vez que un par de gotas entran dentro de las regiones de deformación II y III antes descritas, el programa comienza a contar el tiempo de vida de la película entre ellas. A cada iteración compara el tiempo acumulado ($\tau_{ij}$) con el tiempo característico encontrado por Vrij (1964; 1968) para el mayor crecimiento de ondas superficiales en ausencia de un potencial repulsivo:

$$\tau_{Vrij} = 96\pi^2 \gamma \eta h_0^5 A_H^{-2} \qquad (49)$$

Luego aproxima la amplitud de cada onda superficial por una fracción de la distancia crítica [Urbina-Villalba, 2009a]:

$$\lambda_i = Ran(t) h_c \qquad (50)$$

Donde $Ran(t)$ es un número real aleatorio entre -1.0 y 1.0. En este caso, la coalescencia ocurre si el ancho del film ($h$) es menor o igual a:

$$\lambda_{TOTAL} = (\lambda_i + \lambda_j) \, \exp(\tau_{ij}/\tau_{Vrij}) \qquad (51)$$

### 3.2 Simulaciones de Estabilidad de Emulsiones con Maduración de Ostwald

Justo después de mover las gotas y antes de verificar su posible coalescencia, el programa ejecuta el algoritmo de maduración de Ostwald publicado por De Smet y Finsy [De Smet, 1997]. Partiendo de la ley de Fick, empleando la ecuación de Kelvin y suponiendo que $\alpha \ll R_i$, se puede deducir que el número de moléculas de aceite de una gota ($m_i$) cambia con el tiempo de acuerdo a la siguiente ecuación:

$$\frac{dm_i}{dt} = 4\pi D_m C_\infty \alpha \left( \frac{R_i}{R_c} - 1 \right) \qquad (52)$$

Donde $D_m$ es la constante de difusión en agua de las moléculas de aceite. El último paréntesis se denomina "Ley de Crecimiento", $P_i(t)$, y el término que lo multiplica es la tasa de intercambio de moléculas $M(t)$. Así:

$$m_i \, (t + \Delta t) = m_i \, (t) + M(t) \, P_i(t) \qquad (53)$$

Donde:

$$M(t) = M = 4\pi D_m C(\infty)\alpha \Delta t \qquad m_S \geq M P_S(t)$$
$$M(t) = m_S / M \qquad m_S < M P_S(t) \qquad (54)$$

El subíndice i = "S" se refiere a la gota de menor tamaño existente en la emulsión. Así mientras la gota de menor tamaño posea M moléculas para intercambiar, el intercambio se ejecuta a una tasa constante de intercambio. Sin embargo, cuando la gota tiene un tamaño muy pequeño, solo una fracción de la misma se disuelve.

En un cálculo de ESS sin maduración de Ostwald, el número de partículas típicamente disminuye hasta que todas coalescen a una sola. Sin embargo el proceso de maduración depende del intercambio de aceite entre gotas. Para resolver este problema se implementó una subrutina adicional que cada cierto tiempo aumenta el número de partículas preservando la distribución de tamaño existente en el momento [Urbina-Villalba, 2009]. Si las simulaciones comienzan con un número de partículas igual a: $N(t=0) = N_0$, se deja que el mismo decrezca hasta $N(t=t') = N_0/4$. En ese momento, la caja de simulación originalmente centrada en (x,y,z) = (0,0,0) se traslada con todas sus partículas al cuadrante negativo, mediante un desplazamiento uniforme de todas las gotas existentes en ese instante. Seguidamente, se aplican condiciones de borde periódicas para replicar tres veces la caja original, produciendo una macro-caja que contiene ahora $4(N_0/4) = N_0$ partículas, igual que al principio del cálculo (<span style="color:red">Figura 1</span>). De esta forma, el número de partículas existente nunca disminuye por debajo de un cuarto de las partículas iniciales (ver <span style="color:red">Fig. 1</span>). Por otra parte, las coordenadas relativas de las partículas se preservan entre las partículas de cada una de las cuatro sub-cajas creadas. A partir de esta "regeneración" de la distribución, la caja original es sustituida por la nueva "mega"-caja. El proceso se repite cuantas veces sea necesario de forma automática [Urbina-Villalba, 2009b].

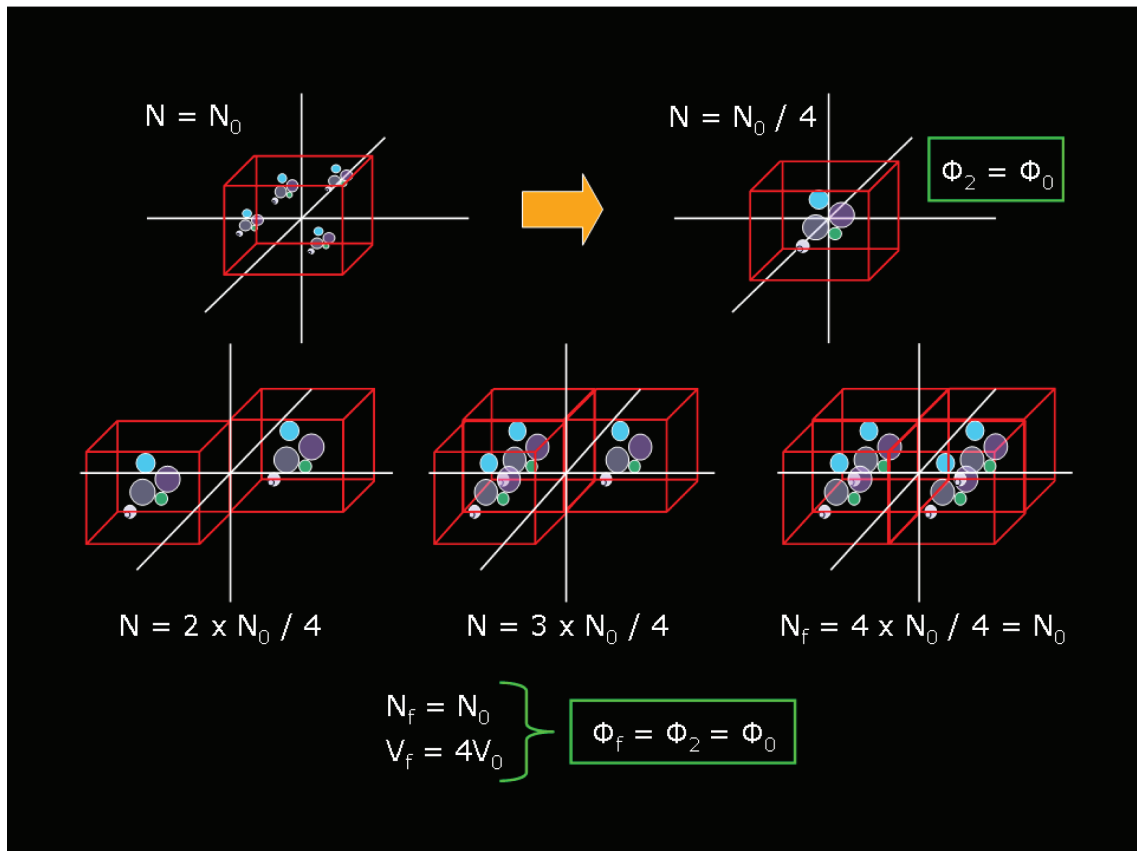

Figura 1: Esquema de Reconstrucción de la Celda de Simulación

## 4. DETALLES COMPUTACIONALES

Todas las simulaciones comienzan a partir de una distribución Gaussiana con un radio promedio de partícula de 30 nm y una desviación estándar de 1.5 nm. Este radio corresponde al menor valor observado por Sakai *et al.* (2002) en sus experimentos. Se emplearon 500 partículas iniciales ($N = N_0 = 500$). El tamaño de la caja de simulación se adaptó para reproducir una fracción de volumen de $\phi = 2.29 \times 10^{-4}$. Tal y como se explicó la Distribución de Tamaño de Gotas (DTG) fue reconstruida cada vez que el número de partículas alcanzaba $N(t) = 125 = N_0/4$, incrementándose el mismo hasta $N = 500$. Se empleó un paso de tiempo de $\Delta t = 1.37 \times 10^{-7}$ s, suficientemente pequeño para muestrear apropiadamente potenciales de corto alcance. El corte en el cálculo del potencial (cut-off) se fijó en 750 nm para tomar en cuenta los potenciales eléctricos correspondientes a fuerzas iónicas bajas. Con éstas aproximaciones los cálculos duraron 2 años en una Dell Precision T7400 de 8 procesadores.

En la ausencia de surfactante la carga superficial de las dispersiones de alcano en agua depende del pH. La adición de NaOH y HCl permite regular la carga superficial pudiéndose obtener variaciones entre -40 mV y -120 mV para pHs entre 5 y 8. En las emulsiones tratadas por Sakai *et al.* (2002), se observó un pH de 6. Por ello, y a menos que se especifique lo contrario, se empleó una fuerza iónica FI de $10^{-6}$ M. Usando este valor, una carga superficial de $\sigma = -0.3$ mCoul/m$^2$ produce un potencial eléctrico superficial de -11.6 mV para una gota de 30 nm. Este se calculó usando la expresión de Sader *et al.* (1995; 1997) implementada en el programa:

$$\sigma_T \varepsilon / \kappa \varepsilon \varepsilon_0 k_B T =$$
$$\Phi_P + \Phi_P / \kappa R_i - \kappa R_i \left(2 \sinh\left(\Phi_P/2\right) - \Phi_P\right)^2 / \overline{Q}$$
$$\overline{Q} = 4 \tanh\left(\Phi_P/4\right) - \Phi_P - \kappa R_i \left[2 \sinh\left(\Phi_P/4\right) - \Phi_P\right]$$
$$(55)$$

Donde: $\Phi_P = \Psi_0 e/k_B T$ es el potencial eléctrico reducido, $\varepsilon_0$ la permitividad del vacío y $\varepsilon$ la constante dieléctrica del agua.

Tabla 1: Parámetros de Cálculo Empleados en las Simulaciones

| Parámetro | Valor |
|---|---|
| $A_H$ (J) | $5.02 \times 10^{-21}$ |
| $C$ ($\infty$) cm$^3$/ cm$^3$ | $5.31 \times 10^{-9}$ |
| $D_m$ (m$^2$/s) | $5.40 \times 10^{-10}$ |
| $V_m$ (m$^3$/mol) | $2.27 \times 10^{-4}$ |
| $\gamma$ (mN/m) | 52.78 |
| $\sigma 1$ (mCoul/m$^2$) | -0.30 |
| $\sigma 2$ (mCoul/m$^2$) | -0.40 |
| $B_0$ (N) | $1.6 \times 10^{-12}$ |
| $\rho$ (g/cm$^3$) | 0.749 |
| $f_0$ (mJ/m$^2$) | 3 |
| $\lambda_0$ (nm) | 0.6 |

Los parámetros de las simulaciones se muestran en la Tabla 1, y algunas características de los potenciales de interacción se muestran en la Tabla 2.

Se estudiaron 7 sistemas los cuales fueron denotados así:

1) $C_{12}$: Emulsión de gotas no-deformables de dodecano en agua con un potencial superficial de $\Psi_0$ = -11.6 mV ($\sigma$ = -0.3 mCoul/m$^2$) apantallado por una fuerza iónica de 0.5 M NaCl. Esta salinidad es suficiente para eliminar completamente el efecto del potencial eléctrico, por lo que la interacción entre dos gotas se reduce a las fuerzas atractivas de van der Waals entre las moléculas de dodecano ($C_{12}$).

2) $C_{12}$-$E_1$: Emulsión de gotas no-deformables de dodecano en agua con un potencial superficial de $\Psi_0$ = -11.6 mV y una fuerza iónica correspondiente a pH = 6 (FI = 10$^{-6}$).

3) $C_{12}$-$E_1$-LN: Equivalente a $C_{12}$-$E_1$ pero empleando una distribución inicial de gotas lognormal en vez de la distribución Gaussiana empleada en el resto de los cálculos.

4) $C_{12}$-$E_2$: Emulsión de gotas no deformables de dodecano en agua con un potencial superficial de $\Psi_0$ = -15.4 mV ($\sigma$ = -0.4 mCoul/m$^2$) y una fuerza iónica FI = 10$^{-6}$.

5) $C_{12}$-$E_1$-H (0.5 M) = $C_{12}$-H: Similar a $C_{12}$ ($\Psi_0$ = -11.6 mV, FI = 0.5 M) pero incluyendo la fuerza de hidratación ($\lambda_0$ = 0.6 nm, $f_0$ = 3 mJ/m$^2$).

6) $C_{12}$-$E_1$-H: Equivalente a $C_{12}$-$E_1$ pero incluyendo el efecto de las fuerzas de hidratación.

7) $C_{12}$-$E_1$-D: Emulsión de gotas _deformables_ de dodecano en agua con un potencial superficial de $\Psi_0$ = -11.6 mV a FI = 10$^{-6}$ M.

Tal y como puede observarse en la Tabla 2, el tamaño de la barrera repulsiva entre dos gotas de dodecano se incrementa en el orden: $C_{12}$ < $C_{12}$-$E_1$ = $C_{12}$-$E_1$-LN < $C_{12}$-$E_2$ < $C_{12}$-H = $C_{12}$-$E_1$-H (0.5 M) < $C_{12}$-$E_1$-H < $C_{12}$-$E_1$-D.

En presencia de una fuerza iónica alta (0.5 M) el potencial eléctrico se apantalla completamente. A fuerza iónica baja (FI = 10$^{-6}$ M) la barrera repulsiva cambia de 4.0 $k_BT$ ($C_{12}$-$E_1$) a 7.1 $k_BT$ ($C_{12}$-$E_2$) cuando la carga superficial se incrementa de $\sigma$ = -0.3 mCoul/m$^2$ a $\sigma$ = -0.4 mCoul/m$^2$. Estos potenciales no exhiben mínimos secundarios dada su larga longitud de decaimiento [Urbina-Villalba, 2009b]. La inclusión del potencial de hidratación produce una barrera considerable de 16.3 $k_BT$ pero esta barrera se sitúa a sólo 0.28 nm de separación entre las gotas, por lo que su efecto es sólo sensible a muy cortas distancias. Por otra parte, la energía necesaria para la deformación de las gotas en ausencia de surfactante es muy grande (~ 77 $k_BT$). Esta barrera es infranqueable si no fuera por la existencia del mecanismo de coalescencia por ondas capilares que se supone que opera en presencia de una película plano-paralela O/W/O.

## 5. RESULTADOS Y DISCUSIÓN

En un trabajo anterior [Urbina-Villalba, 2009b] se encontró que si se parte de una DTG Gaussiana con $R_P$ = 30 nm, y las gotas se fijan en el espacio tal y como presupone la teoría de LSW, el radio promedio de una emulsión dodecano/agua disminuye durante aproximadamente 830 s, tiempo necesario para que la gota de aceite más pequeña desaparezca por disolución. Luego de ese tiempo el radio promedio comienza a aumentar, alcanzando una tasa de $V_{Teórico}$ = 1.2 x $10^{-26}$ m$^3$/s ($r^2$ = 0.9989) para Δt = 0.01082 s (M = 2.04 x $10^{-5}$ s).

Tabla 2: Características de los Potenciales de Interacción Empleados

| Sistema | Posición del Máximo (nm) | Máximo del potencial $V_{max}$ / $k_BT$ | Posición del Mínimo Secundario (nm) | Valor del Mínimo $V_{min}$ / $k_BT$ |
|---|---|---|---|---|
| $C_{12}$ | —— | —— | —— | —— |
| $C_{12}$-$E_1$ = $C_{12}$-$E_1$-LN | 11.3 | 4.05 | —— | —— |
| $C_{12}$-$E_2$ | 9.30 | 7.13 | —— | —— |
| $C_{12}$-$E_1$-H (0.5M) = $C_{12}$-H | 0.29 | 15.9 | 3.53 | -0.36 |
| $C_{12}$-$E_1$-H | 0.29 | 20.0 | —— | —— |
| $C_{12}$-$E_1$-D | 1.84 | 74.3 | 1.89 | 3.08 |

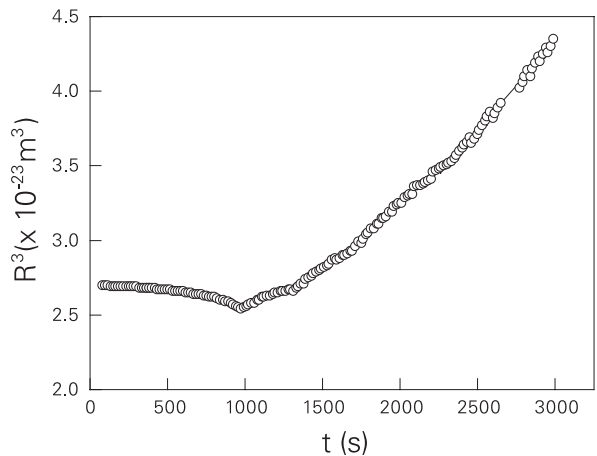

Figura 2: Predicción de $R^3$ vs. t para una emulsión dodecano/
agua (D/W) con partículas fijas en el espacio.

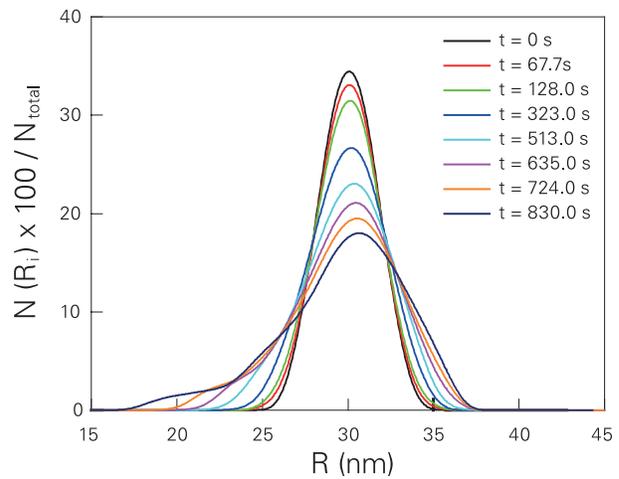

Figura 3: Variación de la DTG en función del tiempo (t ≤ 830
s) para una emulsión D/W con partículas fijas en el espacio
($N_0$ = 500 partículas).

Como demuestra la Fig. 2, el uso de un Δt sustancialmente menor (Δt = 2.2 x $10^{-8}$ s) no cambia los resultados anteriores de manera significativa. El mínimo de la curva puede ubicarse ahora más exactamente alrededor de los 937 s, pero la velocidad de maduración en el régimen estacionario es muy similar $V_{Teórico}$ = 1.05 x $10^{-26}$ m$^3$/s ($r^2$ = 0.99315). Ambos valores son cónsonos con los predichos por la teoría de LSW ($V_{LSW}$ = 1.3 x $10^{-26}$ m$^3$/s) lo cual confirma que el algoritmo de De Smet *et al.* (1997) funciona de acuerdo a lo esperado.

Sin embargo, un análisis detallado de la variación del tamaño promedio en el régimen estacionario demuestra que si las partículas no se mueven, el $R_p$ sólo se incrementa cuando el número de partículas disminuye por disolución de las gotas más pequeñas. Contrario a lo esperado, el radio promedio *siempre disminuye* a consecuencia del intercambio molecular entre gotas cuando no hay disolución de partículas. En consecuencia, la pendiente que se observa en el régimen estacionario es el resultado de dos tendencias opuestas que ocasionan una especie de zig-zag alrededor de la pendiente promedio de la curva de $R^3$ vs. t (ver Fig. 2). Coincidencialmente las medidas experimentales del radio también oscilan en el tiempo respecto al valor promedio, pero tal variación generalmente se adjudica a las propiedades estadísticas de la medida.

Las Figs. 3 y 4 ilustran la evolución de la DTG bajo las mismas condiciones de cálculo (en ausencia de movimiento de las gotas). Obsérvese que la distribución inicial Gaussiana, rápidamente cambia a una DTG del tipo LSW, con cola ha-cia la izquierda y un tamaño máximo de gota que no supera 1.5 veces el tamaño promedio. La disolución progresiva de algunas partículas se evidencia mediante la aparición de pequeños picos que progresivamente se mueven hacia menores radios de partícula hasta que finalmente desaparecen. Eventualmente la distribución degenera completamente si se permite que el número de partículas disminuya de forma indiscriminada hasta que sólo sobrevivan unas pocas.

Si bien las Figs. 3 y 4 reproducen las predicciones de la teoría de LSW, la DTG predicha para un tiempo de 4 minutos (t = 240 s, Fig. 3) difiere sustancialmente de la que se encuentra experimentalmente (ver sección 5.2). En particular, la cola de la señal principal tiende a la derecha en el caso experimental y a la izquierda en el caso teórico.

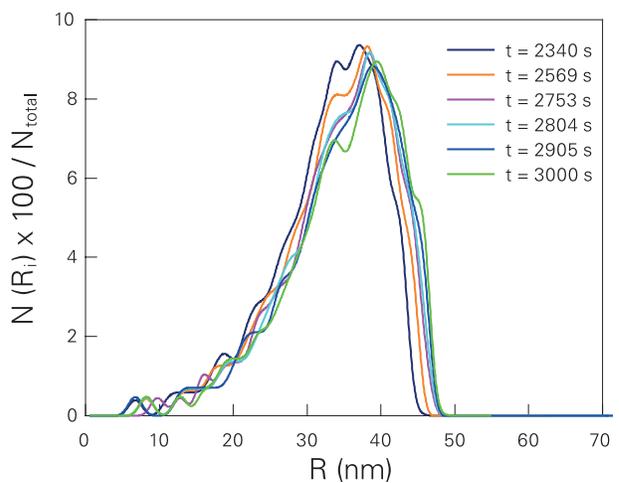

Figura 4: Variación de la DTG en función del tiempo (2340 s ≤ t ≤ 3000 s) para una emulsión D/W con partículas fijas en el espacio ($N_0$ = 500 partículas).

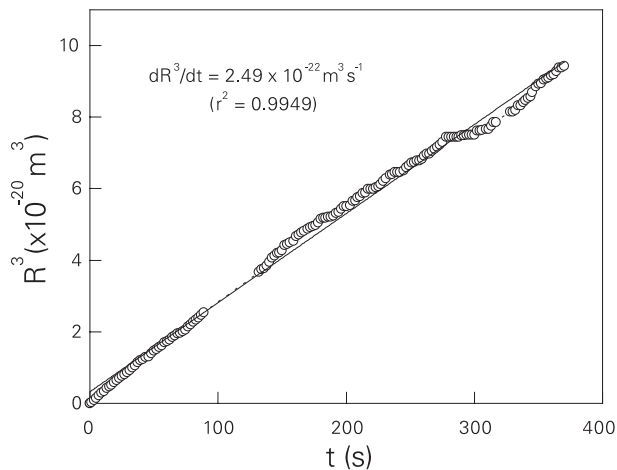

Figura 5: Variación de $R^3$ vs. t para el sistema $C_{12}$ ($C_{12}$-$E_1$ a 0.5 M de NaCl)

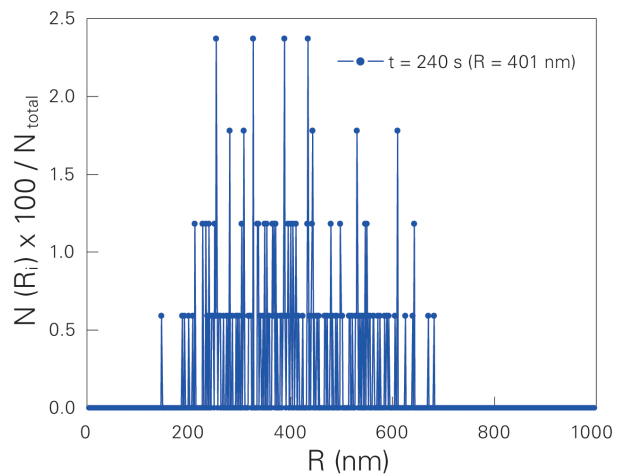

Figura 6: Predicción SEE para la DTG del sistema $C_{12}$

Las Figs. 5 y 6 corresponden al sistema $C_{12}$ e ilustran el resultado de tomar en cuenta el movimiento de las partículas, y su posible coalescencia. En el sistema $C_{12}$ a pesar de que se incluyó un potencial eléctrico entre gotas, el mismo se apantalló con una fuerza iónica elevada, de tal manera que la interacción entre gotas fuese únicamente atractiva, lo cual induce su floculación y coalescencia. De acuerdo a las simulaciones anteriores (Fig. 2) en donde sólo el mecanismo de maduración de Ostwald estaba operativo, no es posible obtener un incremento del radio promedio producto de este fenómeno antes de 900 segundos.

Sin embargo, cuando se permite la agregación y la coalescencia de las partículas se encuentra que la variación temporal del radio cúbico promedio es lineal ($V_{Teórico}$ = 2.5 x 10$^{-22}$ m$^3$/s, $r^2$ = 0.9944) desde el mismo inicio del proceso. Más sorprendente aún: la DTG que se desarrolla producto de la coalescencia de las gotas (no deformables) es totalmente polidispersa, y aún así es compatible con una variación lineal del radio cúbico promedio.

De lo anterior se concluye que:

---

**Conclusión 1**: Los fenómenos de Floculación y Coalescencia pueden dar origen a una variación lineal de $R^3$ vs. t. Tal variación es compatible con una distribución polidispersa de tamaño de gotas.

---

La variación lineal de $R^3$ vs. t anteriormente descrita fue confirmada con medidas experimentales realizadas a tiempos cortos (t < 100 s) sobre suspensiones de partículas de látex a FI = 600 mM [Urbina-Villalba, 2009b]. En condiciones similares (FI = 400 mM y FI = 600 mM), las nanoemulsiones dodecano/agua también presentan una variación temporal aproximadamente lineal, cuya pendiente aumenta con la salinidad. Tal dependencia evidencia la influencia de la floculación en la variación de $R^3$. Sin embargo, el rápido incremento de la polidispersidad de éstos sistemas, y la subsiguiente dispersión múltiple de luz que se origina producto de la agregación, hacen que la tendencia experimental quede completamente enmascarada luego de 100 s, obteniéndose una data que fluctúa de manera aparentemente errática. Por ello no queda claro si las pendientes medidas realmente confirman una genuina variación lineal ó son tangentes a una curva suave. En cualquier caso, la tasa de desestabilización experimental límite (FI = 600 mM) es de $V_{abs}$ = 2.9 x 10$^{-23}$ m$^3$/s ($r^2$ = 0.9517), un orden de magnitud menor que el valor teórico predicho por las simulaciones (2.5 x 10$^{-22}$ m$^3$/s) y que el valor experimental observado en las suspensiones de látex; (1.5 – 1.7) x 10$^{-22}$ m$^3$/s [Urbina-Villalba, 2009].

De lo anterior se deduce que a fuerzas iónicas altas parece existir un potencial repulsivo remanente cuyo efecto no es significativo a salinidades bajas. La mezcla de un potencial de hidratación de corto alcance con un potencial electrostático de largo alcance posee éstas características (Fig. 7). A salinidades bajas (FI = 10$^{-6}$ M) la repulsión electrostática predomina. A salinidades altas, esta repulsión desaparece y sobrevive el potencial de hidratación que opera a muy cortas distancias. Este potencial presenta una repulsión fuerte por debajo de 2 nm, y es levemente atractivo (~ -0.4 $k_B$T) a mayores distancias.

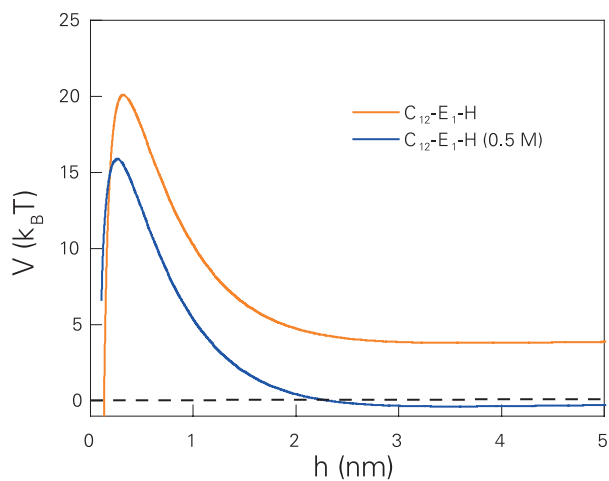

Figura 7: Potencial de interacción para los sistema $C_{12}$-$E_1$-H y $C_{12}$-$E_1$-H (0.5 M).

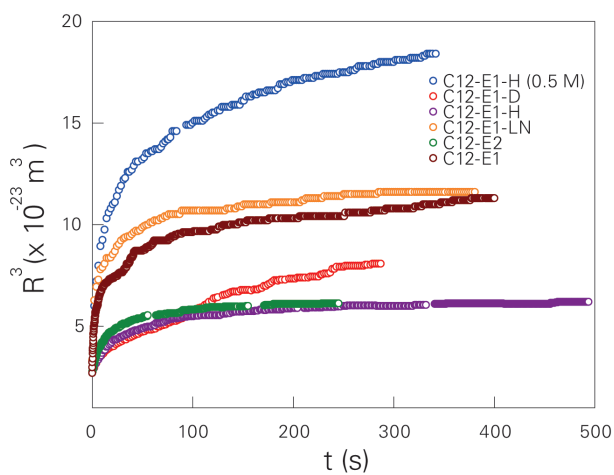

Figura 8: Variación de $R^3$ vs. t para los sistemas estudiados.

Cálculos preliminares para t < 5 segundos incluyendo la fuerza de hidratación (Ec. (48)), arrojaron una pendiente de $1.1 \times 10^{-23}$ m³/s, similar al valor experimental. Esto indujo la realización de simulaciones de mayor extensión. En la Fig. 8 se muestra una de esas simulaciones. Obsérvese que la curva de $R^3$ vs. t correspondiente al sistema $C_{12}$-$E_1$-H (05 M) tiene una pendiente inicial más pronunciada ($\sim 10^{-24}$ m³/s) que la del $C_{12}$-$E_1$, y alcanza asintóticamente valores mayores ($1.0 \times 10^{-25}$ m³/s). Sin embargo, únicamente por debajo de 5 s presenta el orden de magnitud experimental ($10^{-23}$ m³/s). De resto, el potencial repulsivo es demasiado elevado y genera valores de $V_{Teórico}$, sustancialmente menores a los encontrados experimentalmente. Por otra parte debe notarse que la magnitud de $f_0$ en la Ec. (48) es proporcional a la tensión interfacial partícula/agua, por lo que valores mayores a 3 mJ/m² generarían barreras de repulsión superiores. Sin embargo la tensión dodecano/agua es del orden de 50 mN/m². En consecuencia se concluye que la fuerza de hidratación descrita por la Ec. (48) es incapaz de explicar la diferencia entre la teoría y el experimento a fuerzas iónicas altas.

**Conclusión 2**: A salinidades altas existe un potencial repulsivo remanente (no electrostático) y de origen desconocido, que ralentiza la variación de $R^3$ vs. t con respecto al valor límite predicho por las simulaciones. El efecto de este potencial no es significativo a salinidades bajas.

A fin de reproducir la tendencia experimental para FI = 600 mM, se requiere una barrera de potencial de corto alcance y menor tamaño que la barrera electrostática generada por la presencia de cargas superficiales. Sin embargo, ninguno de los potenciales empleados en estos cálculos tiene la forma requerida.

## 5.1 Variación del Radio Cúbico Promedio en una Nano-emulsión Dodecano/Agua

Tales complicaciones y diferencias no ocurren a fuerzas iónicas más bajas, pero en estos casos, la variación de $R^3$ vs. t no es lineal (Fig. 8):

La forma cualitativa de las curvas de $R^3$ vs. t predicha para emulsiones sin surfactante alcano/agua a bajas salinidades (curva cóncava hacia abajo) coincide perfectamente con la forma de las curvas encontradas experimentalmente. Esto se debe en parte a que a medida que las gotas aumentan su tamaño su difusión disminuye, reduciendo su tasa de crecimiento por coalescencia. Por otra parte, si las gotas aumentan de tamaño manteniendo su densidad de carga superficial constante, su carga total debe aumentar, produciendo un incremento del potencial repulsivo electrostático a medida que el tamaño crece. Esto también induciría una reducción de la tasa de agregación y por consiguiente una disminución de la coalescencia.

**Conclusión 3**: La presencia de un potencial repulsivo entre gotas genera una variación no lineal de $R^3$ vs. t. El radio cúbico muestra una curva cóncava hacia abajo cuya pendiente inicial disminuye progresivamente con el tiempo.

Por otra parte, el valor absoluto de la tasa de desestabilización depende del tiempo de medida y del potencial repulsi-vo. Así por ejemplo para 10 s < t < 20 s y FI ~ $10^{-6}$ M (sistema $C_{12}$-$E_1$) el valor de $V_{obs}$ = 1.6 x $10^{-24}$ m³/s, concuerda bastante bien con el valor predicho por las simulaciones: $V_{Teórico}$ = 1.0 x $10^{-24}$ m³/s [Urbina-Villalba, 2009b].

Si tomamos como referencia el sistema $C_{12}$-$E_1$ vemos que el efecto de los demás potenciales repulsivos sobre el valor de $V_{Teórico}$ es el esperado. A medida que la barrera repulsiva es mayor, más lenta es la variación de $R^3$ vs. t (Fig. 8). Así se observa que $V_{Teórico}(C_{12})$ >> $V_{Teórico}(C_{12}$-$E_1)$ > $V_{Teórico}(C_{12}$-$E_2)$ > $V_{Teórico}(C_{12}$-$E_1$-H). En relación al sistema de gotas deformables $V_{Teórico}(C_{12}$-$E_1$-D) debe mencionarse que éstas simulaciones duran muchísimo tiempo de cálculo y que generalmente es necesario reiniciarlas debido a las caídas de electricidad. Esto es intrascendente para la mayoría de los cálculos menos para el de gotas deformables, ya que el mecanismo de coalescencia de Vrij toma en cuenta el tiempo de existencia de la película O/W/O. Dado que no es posible almacenar este tiempo en disco, el mismo se pierde cuando se reinicia una simulación. Esto genera que el tiempo de coalescencia se prolongue artificialmente si los cálculos se caen. Por tanto se espera que $R^3$ vs. t varíe en cada caso de manera más lenta a como lo haría si no hubiese necesidad de reiniciar el cálculos de vez en cuando. Debe resaltarse también que al formarse una película plano-paralela el mecanismo de coalescencia por Vrij se activa, lo cual permite pasar barreras repulsivas que de otro modo son infranqueables. Este mecanismo de coalescencia no esta presente en el resto de los cálculos que corresponden a gotas no-deformables.

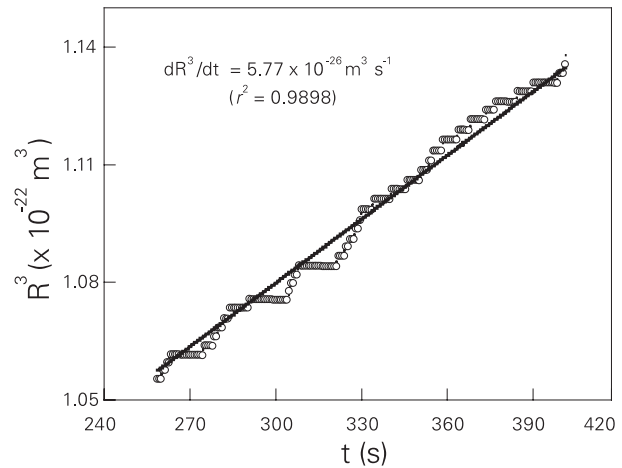

Figura 9: Detalle de la variación de $R^3$ vs. t a tiempos largos para el sistema $C_{12}$-$E_1$.

moviendo el incremento del radio de las partículas pequeñas mediante coalescencia, y previniendo así su posible disolución.

**Conclusión 5**: El incremento de $R^3$ vs. t en emulsiones dodecano/agua durante los primeros 4 minutos luego de su preparación, se debe a los procesos de floculación y coalescencia únicamente.

Si se amplía cualquiera de las curvas de la Figura 8 se hace evidente que el incremento del radio cúbico es el resultado de dos tendencias opuestas (ver Fig. 9). La coalescencia incrementa $R^3$ al disminuir el número de partículas existentes, pero la maduración solamente *DECRECE* el radio promedio a una tasa del orden de $10^{-27}$ m³/s para el caso de dodecano en agua. Es por esta razón que tanto en las simulaciones como en las mediciones experimentales, la data pareciera oscilar alrededor de una pendiente promedio (ver. Fig. 10).

**Conclusión 4**: La tasa $V_{Teórico}$ es menor a medida que el potencial repulsivo entre gotas es mayor, salvo en los casos de gotas deformables o potenciales de muy corto alcance.

La versión actual del programa escribe en un archivo la identificación de las gotas que son eliminadas por disolución, es decir por la transferencia de moléculas de aceite producto del fenómeno de maduración de Ostwald. Sin embargo, ni en la simulación $C_{12}$ ni en ninguna de las 6 simulaciones restantes en donde el mecanismo de Ostwald está activado conjuntamente con el de floculación y coalescencia, se observó la eliminación por disolución. Sólo en el caso en que el alcano es muy soluble en agua (octano) se encuentra tal eliminación. Esto se debe a que si bien las partículas disminuyen su tamaño por intercambio molecular (maduración) las colisiones entre ellas ocurren mucho más rápidamente que este intercambio, pro-

**Conclusión 6**: En las nanoemulsiones alcano/agua en las que el aceite tiene una solubilidad acuosa igual o menor a la del dodecano y la fuerza iónica es suficientemente baja, los fenómenos de floculación y coalescencia previenen la eliminación de gotas por disolución. Es decir, la floculación y la coalescencia incrementan el radio de las gotas a una tasa superior a la que la maduración de Ostwald emplea para disolverlas.

De lo anterior se deriva la conclusión más importante que se puede sacar de las simulaciones de estabilidad de emulsiones con maduración de Ostwald:

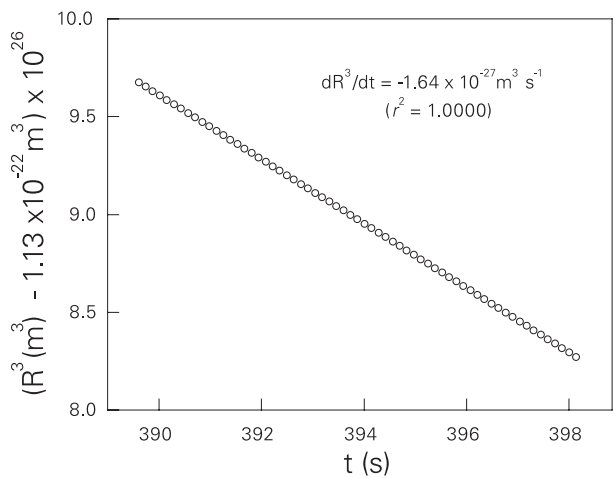

Figura 10: Detalle de la variación de $R^3$ vs. t durante un lapso de tiempo en el que sólo ocurre el intercambio de moléculas de aceite entre gotas sin coalescencia (sistema $C_{12}$-$E_1$).

---

**Conclusión 7**: EL intercambio molecular entre gotas debido a maduración de Ostwald siempre favorece un decrecimiento del radio cúbico *promedio* en función del tiempo. Los valores positivos de $V_{obs}$ obedecen a la disminución del número de partículas bien sea por coalescencia ó por disolución.

---

Nótese que en el caso en que el proceso de floculación está ausente (Fig. 2) también se observa el referido "zig-zag" a tiempos largos, pero en este caso el incremento del radio promedio se debe solamente a la completa disolución de las partículas más pequeñas.

La Tabla 3 muestra los valores asintóticos de las curvas de $R^3$ vs. t mostradas en las Figuras 5 y 8. Nótese que los tiempos totales calculados no son los mismos para todos los sistemas debido a que las emulsiones evolucionan de manera distinta durante el mismo tiempo de cómputo. La Tabla 3 contiene en su segunda columna el valor de $V_{Teórico}$ encontrado hacia el final de la simulación. La tercera columna lista el valor de $V_{Teórico}$ observado durante el último intercambio de moléculas de aceite (maduración de Ostwald) que ocurre en el sistema *sin disminución del número de gotas*.

Excepto por los sistemas $C_{12}$ y $C_{12}$-$E_1$-D, el resto de las dispersiones muestran valores de $V_{Teórico}$ del mismo orden que el valor reportado por Sakai: $V_{obs} = 4.0 \times 10^{-26}$ m$^3$/s [Sakai, 2002]. Cabe re-saltar que el intervalo de tiempo utilizado por Sakai *et al.* (2002) para el cálculo de la tasa de desestabilización fue de más de 100 minutos, mientras que los valores encontrados aquí corresponden a la pendiente de $R^3$ vs. t para 3 min < t < 5 min, por lo que se estima que la concordancia es suficientemente buena. El mayor acuerdo lo exhibe el sistema $C_{12}$-$E_2$ el cual corresponde a gotas de dodecano con un potencial eléctrico superficial de -15.4 mV ($\sigma$ = -0.4 mCoul/m$^2$, FI = $10^{-6}$ M).

El sistema $C_{12}$ discutido anteriormente no presenta ningún potencial repulsivo por lo que su valor de $V_{Teórico}$ es cuatro órdenes de magnitud superior al del resto de los sistemas. Por otra parte, sus lapsos de intercambio molecular sin eliminación de partículas son extremadamente cortos y por lo tanto, la pendiente $V_{Teórico}$ durante el tiempo de eliminación sólo puede calcularse con un máximo de 4 ó 6 datos. Esto causa una incertidumbre muy grande en el valor estimado. En cambio el resto de los datos muestra tasas de intercambio molecular entre -1.1 x $10^{-27}$ y -4.5 x $10^{-27}$ m$^3$/s.

Finalmente cabe destacar que el hecho de que el sistema $C_{12}$-$E_1$-D presente valores de $V_{Teórico}$ distintos al resto de los sistemas y al valor experimental, es una evidencia más de que las gotas de tamaño nanométrico son no-deformables.

### 5.2 Distribución de Tamaño de Gotas de una nano-emulsión Dodecano/Agua a t = 4 minutos.

La Figura 11 muestra la forma cualitativa de la distribución de tamaño de gotas (DTG) de una nanoemulsión de dodecano en agua, 4 minutos después de su preparación [Sakai, 2002]. La distribución es bimodal con una señal amplia de forma lognormal (falda a la derecha) que en la escala

Tabla 3: Valores Asintóticos de $V_{Teórico}$

| Sistema | $V_{Teórico}$ (m$^3$/s) | |
|---|---|---|
| $C_{12}$ | 2.5 x $10^{-22}$ (0 s < t < 370 s) | -2.0 x $10^{-26}$ (333 s < t < 336 s) |
| $C_{12}$-$E_1$ | 5.8 x $10^{-26}$ (260 s < t < 401 s) | -1.6 x $10^{-27}$ (390 s < t < 398 s) |
| $C_{12}$-$E_1$-LN | 5.3 x $10^{-26}$ (76 s < t < 290 s) | -1.4 x $10^{-27}$ (290 s < t < 380 s) |
| $C_{12}$-$E_2$ | 4.1 x $10^{-26}$ (35 s < t < 245 s) | -1.1 x $10^{-27}$ (185 s < t < 219 s) |
| $C_{12}$-$E_1$-H (0.5M) = $C_{12}$-H | 9.7 x $10^{-26}$ (190 s < t < 344 s) | -2.2 x $10^{-27}$ (336 s < t < 344 s) |
| $C_{12}$-$E_1$-H | 2.0 x $10^{-26}$ (45 s < t < 495 s) | -1.6 x $10^{-27}$ (465 s < t < 495 s) |
| $C_{12}$-$E_1$-D | 1.4 x $10^{-25}$ (71 s < t < 288 s) | -4.5 x $10^{-27}$ (279 s < t < 285 s) |

とても高い

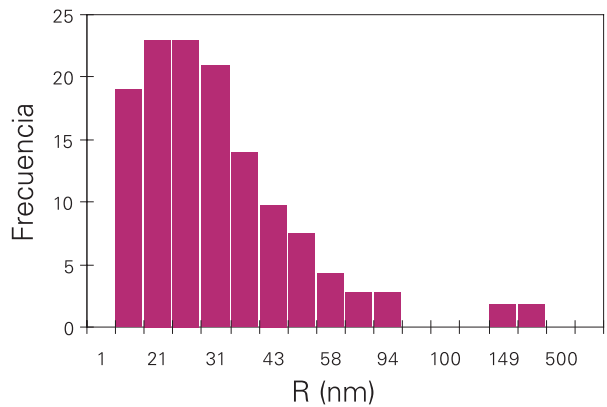

Figura 11: Forma de la Distribución de Tamaño de Gotas (DTG) encontrada por Sakai *et al.* (2002) para una Nanoemulsión de Dodecano en Agua, 4 minutos después de su preparación.

de *radio* de partículas se extiende entre 16 y 94 nanómetros, y una señal adicional separada, de menor intensidad, que cubre entre 149 y 202 nm.

Dada la alta velocidad de desestabilización de las emulsiones de dodecano en agua en ausencia de surfactantes, es imposible medir la DTG de la emulsión inicial (justo después de su preparación). En consecuencia, el estado inicial del sistema, del cual deben partir las simulaciones es desconocido.

Como se explicó con anterioridad, la orientación de la falda de la distribución es una indicación cualitativa del predominio de los procesos de maduración de Ostwald (falda a la izquierda) ó coalescencia (falda a la derecha). En consecuencia y a fin de evitar favorecer un tipo de distribución específico, las simulaciones partieron de distribuciones simétricas de tipo Gaussiano. Sin embargo se realizó un cálculo adicional empleando una DTG inicial lognormal con el fin de evaluar la influencia de la distribución inicial sobre la evolución de la emulsión.

Adicionalmente se supuso que el procedimiento de síntesis por ultrasonido, permitía obtener un tamaño inicial de $R_i = 30$ nm (con una desviación estándar aproximada de 1.5 nm) en el caso del sistema dodecano/agua. Tal conjetura está soportada en las observaciones de Sakai *et al.* (2002) según las cuales 30 nm es el menor tamaño de gota sintetizado por ultrasonido en emulsiones alcano/agua con cadenas hidrocarbonadas entre 6 y 16 carbonos.

Por otra parte debe quedar claro que la forma de la DTG teórica depende de la manera en que las frecuencias del número finito de radios de partícula existentes en la caja de simulación se agrupen para construir el histograma correspondiente. Todos los casos presentados aquí se empleó una resolución de 1.5 nm, es decir 900 categorías de radio distribuidas equitativamente en 1.35 micras. Esa resolución es muchísimo más alta que la que puede proveer ningún aparato experimental. De allí que es sensato esperar que la DTG experimental sea más suave y con un número mucho menor de picos que la DTG calculada.

Debido a las incertidumbres antes descritas, las simulaciones llevadas a cabo únicamente persiguen establecer la evolución de la DTG y su aspecto cualitativo al cabo de 4 minutos.

Sin embargo, a medida que transcurre el tiempo, los cambios de la distribución se hacen más lentos (véanse los cambios entre 53 y 78 s), evidenciando una disminución de la tasa de floculación y coalescencia (FC). Como ya se mencionó tal comportamiento es el resultado de la disminución de la velocidad de difusión de las gotas, y del aumento de su carga superficial total, lo cual retrasa su agregación y ulterior coalescencia. A medida que el proceso FC se retrasa, la maduración de Ostwald gana importancia. Luego de 200 s los cambios se hacen realmente lentos. Véanse por ejemplo las pequeñas variaciones que ocurren entre 240 y 319 s. A medida que este proceso progresa, los picos correspondientes a las partículas de mayor tamaño se hacen mayores, y la falda de la distribución parece cambiar de derecha a izquierda.

De acuerdo a nuestros cálculos los cambios descritos arriba son generales para aquellos aceites de solubilidad igual ó menor que la del dodecano. Sin embargo, la rapidez del proceso depende de las características del potencial de interacción inter-gota. En consecuencia y en ausencia de otros mecanismos de desestabilización, la dirección de la falda de la DTG identifica el predominio del proceso de desestabilización: FC si la falda de la DTG está orientada hacia la derecha ó maduración de Ostwald si la falda de la DTG está orientada hacia la izquierda).

**Conclusión 8**: En una nanoemulsión alcano/agua la dirección en que se desarrolla la falda de la distribución de tamaño de gotas como función del tiempo, identifica el tipo de proceso de desestabilización predominante. Faldas a la derecha señalan la hegemonía del proceso de coalescencia. Faldas a la izquierda indican la preponderancia del fenómeno de maduración de Ostwald.

La Figura 13 muestra las DTG resultantes luego de la evolución de cada uno de los sistemas durante un tiempo de 4 minutos. Como se ilustró anteriormente, los cambios más significativos en la forma de la distribución ocurren durante los primeros segundos. Luego de eso el radio promedio se estabiliza, mostrando valores que van desde 39 nm ($C_{12}$-$E_1$-H) hasta 56 nm ($C_{12}$-$E_1$-H (0.5 M)). El ancho de las distribuciones varía entre 20 y 30 nm, lo cual difiere del ancho la señal principal reportada por Sakai *et al.*

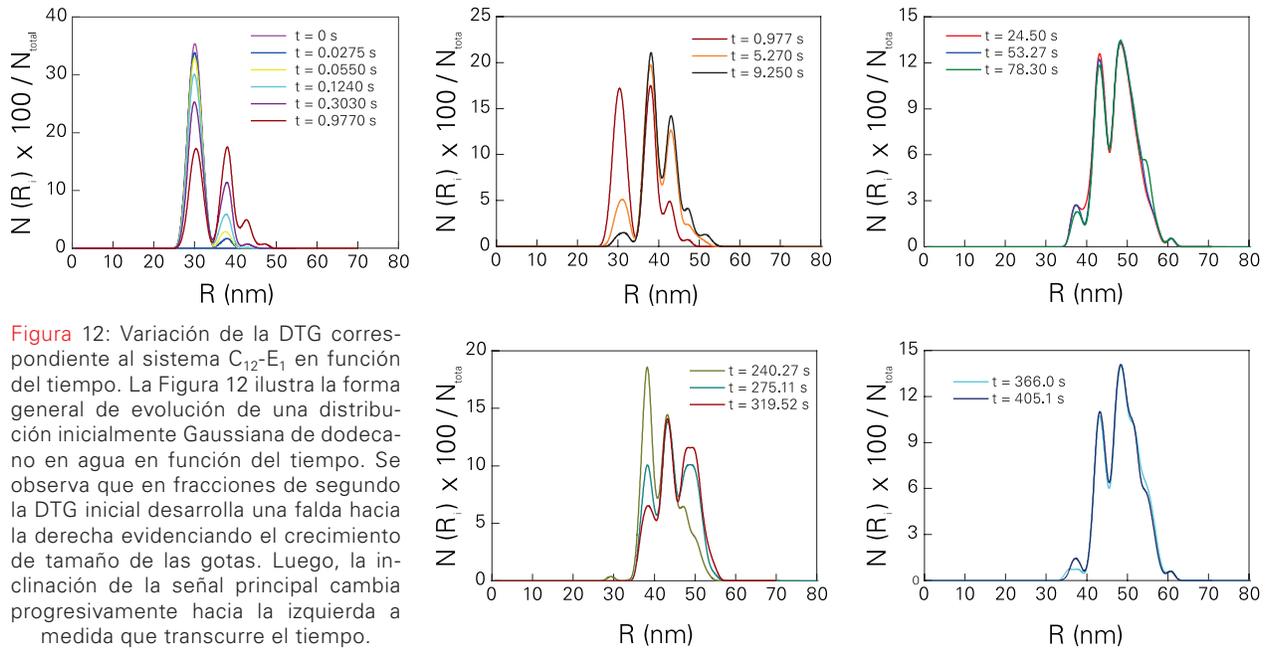

**Figura** 12: Variación de la DTG correspondiente al sistema $C_{12}$-$E_1$ en función del tiempo. La Figura 12 ilustra la forma general de evolución de una distribución inicialmente Gaussiana de dodecano en agua en función del tiempo. Se observa que en fracciones de segundo la DTG inicial desarrolla una falda hacia la derecha evidenciando el crecimiento de tamaño de las gotas. Luego, la inclinación de la señal principal cambia progresivamente hacia la izquierda a medida que transcurre el tiempo.

(2002) para este sistema (78 nm). Tampoco se observa un pico adicional entre 149 y 202 nm como en el experimento. Sin embargo, la comparación de las DTG correspondientes a los sistemas $C_{12}$-$E_1$ y $C_{12}$-$E_1$–LN demuestran que la forma de la distribución de tamaño a los 4 minutos depende marcadamente de la forma de la distribución inicial. De aquí se infiere que las diferencias entre la teoría y el experimento pueden deberse a que la DTG inicial experimental es mucho más polidispersa que la distribución Gaussiana empleada en estos cálculos.

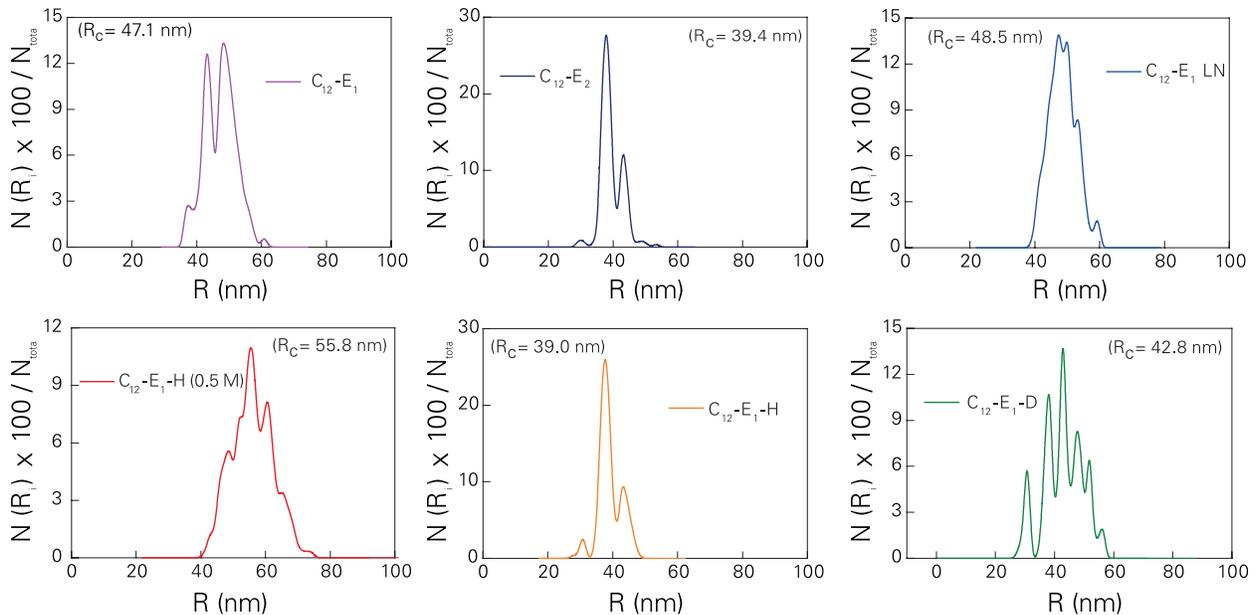

Figura 13: Distribuciones de Tamaño de Gotas para t = 4 minutos.

## 6. CONCLUSION GENERAL

Los resultados anteriores demuestran que a pesar de la sofisticación de las simulaciones realizadas, la evolución de una simple nanoemulsión aceite/agua en ausencia de surfactante es difícil de predecir. Esto se debe fundamentalmente al desconocimiento de la DTG inicial, y a los rápidos cambios que ocurren sobre esta durante los primeros segundos. A pesar de estas limitaciones, el valor de $V_{obs}$ si puede justificarse fácilmente a partir de las simulaciones. Curiosamente el valor obtenido resulta fundamentalmente de los procesos de floculación y coalescencia.

## 7. AGRADECIMIENTOS

## 8. BIBLIOGRAFÍA